\documentclass[11pt]{article}

\usepackage[utf8]{inputenc}
\usepackage[T1]{fontenc}
\usepackage[margin=1in]{geometry}
\usepackage{times}
\usepackage{booktabs}
\usepackage{hyperref}
\usepackage{xcolor}
\usepackage{enumitem}
\usepackage{tabularx}
\usepackage{listings}
\usepackage{natbib}
\usepackage{amsmath}
\usepackage{amssymb}
\usepackage{float}
\usepackage{graphicx}
\usepackage{subcaption}
\usepackage{orcidlink}
\usepackage{microtype}
\usepackage{tikz}
\usetikzlibrary{arrows.meta,positioning,calc,fit,backgrounds,decorations.pathreplacing}

\hypersetup{
    colorlinks=true,
    linkcolor=blue,
    citecolor=blue,
    urlcolor=blue
}

\title{sciwrite-lint: Verification Infrastructure\\for the Age of Science Vibe-Writing}

\author{
    Sergey V. Samsonau\,\orcidlink{0000-0002-0835-2970}\thanks{Corresponding author: \href{mailto:ssamsonau@gradcenter.cuny.edu}{ssamsonau@gradcenter.cuny.edu}}\\[6pt]
    Authentic Research Partners, Princeton, NJ\thanks{\url{https://arpconnect.com/}}
}

\date{May 2026}

\begin{document}
\pagestyle{plain}
\maketitle


\begin{abstract}
Scientific papers make claims about prior work backed by citations. Verifying those citations at scale (that each cited paper exists, says what the citation claims, and is itself reliable) is structurally beyond what human review can deliver: a typical paper has dozens of citations, and a careful reviewer reads at most a handful end-to-end. AI-assisted writing makes this gap even more urgent: LLMs hallucinate references and may fill in plausible details from titles or abstracts of papers they never read, worse for the smaller local-weights models that privacy-aware researchers must use.

\textbf{sciwrite-lint}\footnote{Source: \url{https://github.com/authentic-research-partners/sciwrite-lint}. PyPI: \url{https://pypi.org/project/sciwrite-lint/}.} applies the \emph{linting paradigm} from software engineering to citation verification: it runs entirely on the researcher's machine (free public databases, a single consumer GPU, and open-weights models), is fast enough to re-lint between revisions so authors catch problems at the source while drafting, and serves journals and reviewers as an automated first pass. The pipeline checks reference existence, metadata accuracy, retraction status, and claim support, traverses one level into cited papers' bibliographies, and produces per-reference reliability scores. We evaluate on 30 unseen papers (arXiv and bioRxiv) with error injection and LLM-adjudicated false-positive analysis.

The same linting workflow extends to internal consistency: numbers in text vs.\ tables, abstract vs.\ body, figure captions vs.\ content, statistical results vs.\ their verbal interpretation, plus structural cross-references (dangling cites, orphan references). As a separate experimental contribution we also propose SciLint Score: citation-chain integrity combined with a contribution component operationalizing five philosophy-of-science frameworks (Popper, Lakatos, Kitcher, Laudan, Mayo).
\end{abstract}


\section{Introduction}
\label{sec:intro}

A typical scientific paper makes claims about prior work in the scientific community and backs them up with citations. Evaluating those claims requires checking, for each citation, that the cited paper exists, says what the citation claims it says, and is itself reliable.

AI-assisted writing makes citation verification more important than ever. \emph{Science vibe-writing} (accepting AI-generated content because it reads fluently, without systematic verification) produces hallucinated references, fabricated quotes, and misattributed claims. A separate problem: even when an LLM cites a real paper, it may not have actually read the source. Many journals are paywalled, so the model may have seen only the abstract during training; as publishers move to gate even abstracts behind login walls, the model may have only the title to work from. In either case the model may fill in ``plausible details'' about the cited paper's content without telling the user it never read the source. The problem compounds with local open-weights models, which hallucinate more than frontier cloud ones \citep{xu2026ghostcite}; yet local models are exactly what privacy-conscious researchers must use, because cloud-based AI exposes unpublished manuscripts to third-party infrastructure (in March 2023, Samsung's semiconductor division identified three incidents in three weeks of engineers pasting proprietary source code and internal meeting notes into ChatGPT, leading to a company-wide ban on generative AI \citep{samsung2023chatgpt}), training data can be extracted from production language models \citep{nasr2023extracting}, and major ML conferences (NeurIPS, ICML, ACL) now prohibit reviewers from uploading confidential submissions to cloud AI services. A verification tool used in this setting must therefore also run locally.

Existing quality-assurance systems do not catch citation errors at scale. Peer review samples a few citations at best, and even at top venues misses fabricated citations entirely: GPTZero found 100 fabricated citations in NeurIPS 2025 papers, each of which had passed review by 3--5 expert reviewers \citep{ansari2026compound}; all four HPC conferences analyzed by \citet{bienz2026mysterious} required AI disclosure, with no author complying. Preprint servers and open repositories have only moderation and no detailed review process: the only substantive filter between a manuscript and the public scientific record is the author's integrity and the time they have available to verify, the latter constrained by the ``publish or perish'' phenomenon. What is missing is tooling that operates on the paper itself: automatically, at scale, and fast enough to give the author feedback while drafting rather than after submission. This is the \emph{linting paradigm} from software engineering, where editor-level checks catch bugs at the source rather than at code review. \textbf{sciwrite-lint} applies this paradigm to scientific manuscripts. It verifies references, checks claim support, and traces evidence through the citation graph one level deep into cited papers' own bibliographies, all on the researcher's machine with no manuscript content sent to external services. Because it runs locally and is fast enough to re-lint after each revision, an author working in a local vibe-writing workflow can catch fabricated references, misattributed claims, and unsupported statements while the text is still in the draft. The same tool serves the reviewer side: journals, conferences, and individual reviewers can run sciwrite-lint as an automated first pass on submitted manuscripts, or use it semi-automatically alongside human review.

The same linting workflow extends beyond reference verification to internal consistency. Manuscripts contain dense cross-references among abstract, body, tables, and figures, and these can drift independently of any citation problem: an AI-assisted edit changes a number, label, or claim in one place but not in others, and the author no longer notices on re-read. sciwrite-lint surfaces such drift across the manuscript (numbers in text vs.\ tables, abstract claims vs.\ body, figure captions vs.\ content, axis labels vs.\ text, statistical results vs.\ their verbal interpretation), alongside structural cross-reference checks (dangling cites, orphan references, unreferenced figures). The full inventory of checks appears in Table~\ref{tab:checks}.

Beyond citation and internal-consistency verification, the broader research-assessment system also misses the mark on \emph{contribution assessment}: citation counts, journal rank, and the h-index track attention rather than the substance of what a paper contributes. The Matthew Effect concentrates citations regardless of intrinsic paper quality: identical duplicate papers in higher-impact-factor journals receive on average twice the citations of their counterparts in lower-impact-factor journals \citep{lariviere2010matthew}. DORA, signed by 20,000+ researchers, calls for abandoning journal impact factor in hiring and funding \citep{dora2012}; the HEFCE-commissioned Metric Tide review proposes a framework for responsible use of research metrics \citep{wilsdon2015metrictide}. Perverse incentives corrupt integrity \citep{edwards2017perverse}. Journal rank poorly predicts research quality \citep{brembs2013deep}. The h-index is mathematically inconsistent \citep{waltman2012hindex}. As a separate, experimental contribution we also propose \emph{SciLint Score} (Section~\ref{sec:scilint}): citation-chain integrity combined with a contribution component drawn from philosophy of science.

\subsection{Contributions}

\begin{enumerate}
    \item \textbf{sciwrite-lint}\footnote{Version 0.5.0 at the time of publication. Source: \url{https://github.com/authentic-research-partners/sciwrite-lint}. PyPI: \url{https://pypi.org/project/sciwrite-lint/}.}: open-source verification pipeline, fully local, available on PyPI, running on free databases and open-weights models (Section~\ref{sec:chain}).
    \item \textbf{Automatic bibliography verification}: the first system that traces claims through the citation graph by automatically downloading, parsing, and verifying cited papers, including their own bibliographies (Section~\ref{sec:chain}).
    \item \textbf{Structured pipeline over agentic architecture}: deterministic checks where possible, LLM as semantic engine where language understanding is required. Reproducible and auditable on a single consumer GPU (Section~\ref{sec:chain}).
    \item \textbf{Real-world corpus evaluation}: error injection with 98.5\% recall, plus LLM-adjudicated false positive analysis on 30 unseen papers (Section~\ref{sec:realworld}).
    \item \textbf{SciLint Score (open proposal)}: a proposed metric combining integrity verification with contribution assessment from philosophy of science, released as experimental code (Section~\ref{sec:contribution}).
\end{enumerate}

Table~\ref{tab:tools} positions sciwrite-lint against existing verification tools across the same scope and accessibility dimensions; Section~\ref{sec:related} discusses each tool in detail.

\begin{table}[ht]
\centering
\footnotesize
\begin{tabularx}{\textwidth}{Xccccccc}
\toprule
& \rotatebox{55}{\textbf{CiteAudit}} & \rotatebox{55}{\textbf{CiteVerifier}} & \rotatebox{55}{\textbf{RefChecker}} & \rotatebox{55}{\textbf{SemanticCite}} & \rotatebox{55}{\textbf{Publisher}$^\dagger$} & \rotatebox{55}{\textbf{Commercial}$^\ddagger$} & \rotatebox{55}{\textbf{sciwrite-lint}} \\
\midrule
\textbf{Scope} & & & & & & & \\
\quad Ref.\ existence         & \checkmark & \checkmark & \checkmark &            & \checkmark & \checkmark & \checkmark \\
\quad Metadata accuracy       & \checkmark &            & \checkmark &            &            & \checkmark & \checkmark \\
\quad Matching engine$^{**}$  &            &            &            &            &            &            & \checkmark \\
\quad Retraction detection    &            &            &            &            & \checkmark &            & \checkmark \\
\quad Claim--citation support &            &            &            & \checkmark &            & \checkmark & \checkmark \\
\quad Figure--text verification &            &            &            &            &            &            & \checkmark \\
\quad Internal consistency    &            &            &            &            &            &            & \checkmark \\
\quad Bib.\ verification      &            &            &            &            &            &            & \checkmark \\
\quad In-depth bib.\ check  &            &            &            &            &            &            & \checkmark \\
\quad Citation purpose$^*$    &            &            &            &            &            &            & \checkmark \\
\quad AI-generation detection &            &            &            &            & \checkmark & \checkmark &            \\
\quad SciLint Score           &            &            &            &            &            &            & \checkmark \\
\midrule
\textbf{Infrastructure} & & & & & & & \\
\quad Open-source        & ?          & \checkmark & \checkmark & \checkmark &            &            & \checkmark \\
\quad \texttt{pip install}       &            &            & \checkmark &            &            &            & \checkmark \\
\quad Manuscript local$^\S$  & ?          & \checkmark & \checkmark & \checkmark   &            &            & \checkmark \\
\quad Auditable data flow   & ?          & \checkmark & \checkmark & \checkmark   &            &            & \checkmark \\
\quad Runs on    & GPU cluster & CPU        & CPU        & CPU          & Service    & Service    & Consumer GPU \\
\bottomrule
\end{tabularx}
\caption{Verification tools compared. ? = claimed but no code released. $^*$Springer Nature performs binary relevance filtering; sciwrite-lint classifies into eight roles with graduated weights. $^{**}$Multi-signal scoring for papers without DOIs (Section~\ref{sec:matching-engine}). $^\S$Manuscript text never sent externally. $^\dagger$Publisher: Problematic Paper Screener, STM Integrity Hub, Springer Nature Irrelevant Reference Checker. $^\ddagger$Commercial: scite.ai, GPTZero. Figure--text verification is semantic (caption matches content, text describes figure accurately); distinct from image forensics (duplication/manipulation detection by Proofig, Imagetwin, STM Hub's Snappshot).}
\label{tab:tools}
\end{table}


\section{Why and how scientists use citations}
\label{sec:citation}

Section~\ref{sec:intro} described what citation verification needs to do at scale. This section steps back to more basic questions: why do scientists cite at all, how do they cite in practice, and where does that practice diverge from the normative ideal?

The traditional normative view, dominant in mid-twentieth-century sociology of science, frames citation as an honest record of intellectual debt: under this view, citation counts approximate true intellectual impact. Decades of competing accounts tell a different story: citations function as rhetorical instruments that persuade readers, as credibility supports recruited to fortify claims, and as shorthand pointers to ideas invoked without reading the source. \citet{tahamtan2019citingbehavior}, reviewing two decades of empirical work on citing behavior, conclude that citation decisions are multidimensional and shaped by many scientific and non-scientific factors.

Empirical studies confirm this picture. \citet{simkin2003read} analyzed misprint propagation and found 70--90\% of citations are copied from other reference lists without reading the original. \citet{mogull2017quotation}, reviewing 15 studies, recalculated the rate of quotation errors at 14.5\% (95\% CI 10.5--18.6\%); most are major errors where the cited source fails to substantiate, contradicts, or is unrelated to the assertion. Reference lists are inflating: 29 per paper in 2003, 45 in 2019 \citep{dai2021references}.

Generative AI amplifies these same patterns at machine scale. LLMs hallucinate references at rates of 14--95\% across 13 models tested \citep{xu2026ghostcite}, from DeepSeek (14\%) to Hunyuan (95\%). GhostCite documented the same fabricated citation in 16 independent papers. But the core failure is the same: citations included without reading the cited work, without verifying the claim, and without an argumentative reason for the citation's presence. Beyond fabrication, frontier LLMs silently corrupt documents during delegated edits: across 52 long-document domains, even Gemini 3.1 Pro, Claude 4.6 Opus, and GPT 5.4 corrupt an average of 25\% of document content over chained edit relays, with errors that are sparse, severe, and compound across rounds \citep{laban2026delegate}.

The problem compounds through the citation graph. When an AI-assisted paper cites another AI-assisted paper, fabricated references propagate machine-to-machine with no human verification at any level. A separate failure mode is invisible even to careful readers: LLMs have no access to retraction status at generation time, so a paper retracted for fabrication is cited as readily as any other. Detection requires cross-referencing each cited DOI against the Retraction Watch database (60,000+ entries), a check no human author routinely performs.

\subsection{When is a citation justified?}

Across five frameworks from philosophy of science, the answer converges: \textbf{a citation is justified if and only if removing it would weaken the argument}. Each framework reveals a specific argumentative function:

\begin{table}[ht]
\centering
\small
\begin{tabularx}{\textwidth}{llXr}
\toprule
\textbf{Function} & \textbf{Framework} & \textbf{What it does} & $w_i$ \\
\midrule
Evidence & Popper & Supports a specific falsifiable claim & 1.0 \\
Contrast & Mayo & Tested against, improved upon, disagreed with & 0.9 \\
Method & Laudan & Methodology or technique provenance & 0.8 \\
Definition & --- & Establishes terminology & 0.7 \\
Example & --- & Illustrative instance of a broader point & 0.6 \\
Attribution & Laudan & Credits origin of an idea or concept & 0.5 \\
Tool & --- & Tool or dataset provenance & 0.4 \\
\midrule
Context & --- & Background without a specific claim & 0.2 \\
\bottomrule
\end{tabularx}
\caption{Citation functions and their weights in the SciLint Score integrity equation ($w_i$ in Equation~\ref{eq:scilint}). Functions above the line serve the argument; \emph{context} does not.}
\label{tab:cite_functions}
\end{table}

The weights encode a principled hierarchy: an evidence citation that fails verification ($w_i = 1.0 \times 0.0$) is a serious integrity problem; a context citation that fails ($w_i = 0.2 \times 0.0$) barely registers, because it wasn't doing argumentative work. A paper where every citation is classified as \emph{context} has low integrity weight even if all references exist; it is citing without arguing. \emph{Example} is distinct from \emph{evidence}: an illustrative citation does not claim the source supports the broader argument. The source only needs to \emph{be} the instance described.

Decades of attempts to reform citation culture have not closed these gaps. Automation is the remaining lever: verify that the citation chain is intact and that each citation earns its place in the argument. Several tools attempt parts of this (Table~\ref{tab:tools}); sciwrite-lint is distinguished by combining broad scope (existence, metadata, retraction, claim support, bibliography traversal, internal consistency, figure--text alignment, citation purpose) with broad accessibility (open-source, \texttt{pip install}-able, runs locally on a consumer GPU).


\section{The Verification Pipeline}
\label{sec:chain}

Checking that a reference exists in CrossRef is necessary but insufficient. The cited paper must actually support the claim, and the evidence \emph{it} rests on must hold up in turn. Verification must trace through the citation graph automatically.

\subsection{The Pipeline}

Given a manuscript (LaTeX or PDF), the pipeline executes:

\begin{enumerate}[nosep]
    \item \textbf{Figure analysis + text checks + LLM consistency} (no network): a vision-language model (Qwen3-VL, 2B or 8B, user-selectable) extracts structured descriptions from all figures in the manuscript (LaTeX or PDF input). Descriptions are cached by content hash. Then: dangling citations, dangling cross-references, pairwise section contradictions, structural promises, and eleven full-paper checks (Table~\ref{tab:checks}): seven numerical/consistency checks plus four figure checks that cross-reference visual content against text and captions. All full-paper checks share a single cached prefix (paper body + figure descriptions) via vLLM's automatic prefix caching (APC). Runs \emph{concurrently} with step 2.
    \item \textbf{Verify references} (network): structured identifiers (DOI, arXiv ID, PMID, ISBN, LCCN) are checked via deterministic API calls. When no identifiers are available, the matching engine scores candidates across title, author, year, and venue (Section~\ref{sec:matching-engine}). Metadata is compared against canonical records. DOIs are cross-referenced against Retraction Watch (${\sim}$60K entries, cached locally). A cross-ID validation step checks for conflicting identifiers within each entry. LLMs routinely mix metadata from multiple papers, producing entries where the DOI resolves to one paper and the arXiv ID to another.
    \item \textbf{Download + parse}: user-provided local PDFs are matched first (citekey-prefix filenames bind authoritatively; remaining filenames go through fuzzy title matching at threshold 0.80). Remaining papers are downloaded from 14 open-access sources. Authors can also drop web captures (Markdown or MHTML) into a parallel \texttt{local\_web\_dir} folder; MHTML is converted to Markdown at ingest, and pages backing footnote \texttt{\textbackslash url\{...\}} citations participate in claim verification. All PDFs are parsed via GROBID into section-structured markdown and embedded (Snowflake Arctic Embed M v2.0) for semantic retrieval. GROBID also extracts each cited paper's own bibliography into structured entries. Every fetched PDF passes a multi-signal validation gate against the bibliography evidence (title similarity, first-page surname / DOI / year, landing-page detection); a DOI match alone is necessary but not sufficient, so a wrong DOI on an otherwise plausible bib entry can no longer silently land the wrong paper in the workspace.
    \item \textbf{Check cited papers' consistency} (GPU, fast): raster figures are extracted from each cited paper's PDF and described by the VL model (batched across all papers). Then: pairwise section contradictions, structural promises, and eleven full-paper checks (with figure descriptions) within each cited paper. Two concurrent batches at different reasoning depths. GPU stages are sequential: the VL model loads and unloads before vLLM queries begin. Runs \emph{before} claim verification to avoid memory contention.
    \item \textbf{Verify claims + bibliographies} (GPU + network, concurrent):
    \begin{itemize}[nosep]
        \item \textbf{Verify claims} (GPU): each claim is embedded and matched against the cited paper's sections via KNN retrieval. The LLM receives only the relevant sections alongside the claim. If not supported, a context-narrowing step extracts the specific sentence(s), fuzzy-matches them back to the cited paper's text, and re-verifies.
        \item \textbf{Verify cited papers' bibliographies} (network): batch-check existence and metadata of each cited paper's own references via OpenAlex and Semantic Scholar. Produces per-reference hallucination rate and metadata mismatch count.
    \end{itemize}
    \item \textbf{Aggregate}: per-reference reliability score combining all signals (metadata, consistency, claims, bibliography). The \texttt{reference-unreliable} check fires when convergent evidence flags a reference.
\end{enumerate}

\begin{figure}[H]
\centering
\footnotesize
\begin{tikzpicture}[
    >=Stealth,
    box/.style={draw, rounded corners=1.5pt, minimum height=0.45cm,
                minimum width=1.3cm, font=\scriptsize, align=center, inner sep=2pt},
    op/.style={box, fill=violet!8},
    opfut/.style={box, densely dashed, draw=gray!50, fill=violet!4, text=gray!50},
    paper/.style={draw, rounded corners=2pt, fill=blue!8, minimum height=0.5cm,
                  minimum width=1.2cm, font=\scriptsize, align=center, inner sep=2pt},
    paperfut/.style={paper, densely dashed, draw=gray!50, fill=blue!4, text=gray!50},
    arr/.style={->, semithick},
    arrf/.style={->, semithick, densely dashed, gray!50},
    treearr/.style={-, semithick},
    lbl/.style={font=\scriptsize\bfseries},
]

\node[font=\small\bfseries] at (3.5, 3.8) {Per-paper operations};

\node[op] (fa) at (0, 3.0) {Figure\\[-2pt]analysis};
\node[op, right=0.2cm of fa] (cc) {Consistency\\[-2pt]checks};
\node[op, densely dashed, right=0.2cm of cc] (cs) {Contribution\\[-2pt]scoring};
\draw[arr] (fa)--(cc);

\node[op] (s1) at (0, 1.9) {Identify\\[-2pt]references};
\node[op, right=0.3cm of s1] (s2) {Verify\\[-2pt]existence};
\node[op, right=0.3cm of s2] (s3) {Download\\[-2pt]full text};
\node[op, right=0.3cm of s3] (s4) {Parse +\\[-2pt]embed};
\node[op, right=0.3cm of s4] (s5) {Verify\\[-2pt]claims};
\draw[arr] (s1)--(s2); \draw[arr] (s2)--(s3);
\draw[arr] (s3)--(s4); \draw[arr] (s4)--(s5);

\begin{scope}[on background layer]
\node[draw=gray!40, rounded corners=4pt, fill=gray!4,
      fit=(cc)(cs)(s1)(s5), inner sep=6pt, inner ysep=8pt] (pipeline) {};
\end{scope}


\node[paper] (p0) at (2.5, -0.2) {Manuscript};
\node[font=\tiny, align=left, text=black!70, right=0.3cm of p0]
    {all operations};

\node[paper] (r1a) at (0.5, -1.4) {Ref A};
\node[paper] (r1b) at (2.5, -1.4) {Ref B};
\node[paper] (r1c) at (4.5, -1.4) {Ref C};
\node[font=\scriptsize] at (5.8, -1.4) {\ldots};
\node[font=\tiny, align=left, text=black!70] at (6.5, -1.4)
    {downloaded, parsed,\\[-1pt]self-checked; claims\\[-1pt]not verified};

\draw[treearr] (p0.south) -- ++(0,-0.3) -| (r1a.north);
\draw[treearr] (p0.south) -- ++(0,-0.3) -| (r1b.north);
\draw[treearr] (p0.south) -- ++(0,-0.3) -| (r1c.north);

\node[draw, densely dashed, rounded corners=2pt, fill=blue!8,
      minimum height=0.5cm, minimum width=1.2cm, font=\scriptsize,
      align=center, inner sep=2pt] (r2a1) at (0.0, -2.6) {Ref A1};
\node[draw, densely dashed, rounded corners=2pt, fill=blue!8,
      minimum height=0.5cm, minimum width=1.2cm, font=\scriptsize,
      align=center, inner sep=2pt] (r2a2) at (1.5, -2.6) {Ref A2};
\node[draw, densely dashed, rounded corners=2pt, fill=blue!8,
      minimum height=0.5cm, minimum width=1.2cm, font=\scriptsize,
      align=center, inner sep=2pt] (r2b1) at (3.0, -2.6) {Ref B1};
\node[font=\scriptsize] at (4.3, -2.6) {\ldots};
\node[font=\tiny, align=left, text=black!70] at (6.5, -2.6)
    {existence + metadata\\[-1pt]checked (not downloaded)};

\draw[treearr, densely dashed] (r1a.south) -- ++(0,-0.25) -| (r2a1.north);
\draw[treearr, densely dashed] (r1a.south) -- ++(0,-0.25) -| (r2a2.north);
\draw[treearr, densely dashed] (r1b.south) -- ++(0,-0.25) -| (r2b1.north);

\node[draw, rounded corners=3pt, fill=green!4,
      font=\small\bfseries, minimum height=0.6cm,
      minimum width=3cm, align=center, inner sep=4pt]
    (scilint) at (3.0, -3.7) {Verification report};

\end{tikzpicture}
\caption{Verification architecture. \textbf{Top:} per-paper operations (steps 1--6 above). \textbf{Bottom:} the pipeline fans out through the citation graph at three levels: the manuscript (all operations), cited papers (downloaded, parsed, consistency-checked), and their references (existence + metadata checked via API). Solid borders: full text available; dashed: API-verified only.}
\label{fig:depths}
\end{figure}

Figure~\ref{fig:depths} illustrates the architecture. The pipeline follows citations one level deep: cited papers are downloaded, parsed, consistency-checked, and their bibliographies verified via API. Full text of those references is not downloaded, as that would require a second layer of parsing and embedding. The architecture supports deeper traversal by repeating the same stages; as consumer GPUs scale, this becomes practical without architectural changes.

For a typical paper with 50 references (30 with downloadable full text), \texttt{sciwrite-lint check} requires ${\sim}$800 LLM calls; embedding pre-filtering reduces the total from ${\sim}$2,000 to ${\sim}$800 by selecting the most relevant sections per claim. An initial run takes up to 30 minutes (dominated by claim verification); cached runs complete in a few minutes or less. \texttt{sciwrite-lint contributions} adds ${\sim}$400 calls.

\subsection{Reference Integrity Scoring}

The pipeline estimates reference integrity $S(r_i)$ from metadata signals collected during verification:

\begin{table}[ht]
\centering
\small
\begin{tabularx}{\textwidth}{Xr}
\toprule
\textbf{Signal} & \textbf{Score} \\
\midrule
T1 (API-verified + full text) & 0.9 \\
T2 (API-verified, no full text) & 0.7 \\
T3 (not found in APIs) & 0.3 \\
Retracted & 0.0 \\
Expression of Concern (multiplier) & $\times 0.3$ \\
Per metadata mismatch (title/author/year/venue) & $-0.1$ each \\
Per cross-ID mismatch (DOI vs arXiv vs PMID) & $-0.1$ each \\
Non-formal document (news, guide, etc.) & $-0.2$ \\
\midrule
\multicolumn{2}{l}{\emph{From LLM consistency checks (step 4)}} \\
Per consistency warning in cited paper & $-0.05$ \\
Per consistency error in cited paper & $-0.10$ \\
\midrule
\multicolumn{2}{l}{\emph{From bibliography verification (step 5)}} \\
Bibliography hallucination rate & proportional \\
Per bib.\ metadata mismatch (capped $-0.30$) & $-0.05$ \\
Per bib.\ retraction found (capped $-0.30$) & $-0.15$ \\
\bottomrule
\end{tabularx}
\caption{Reference reliability signals. Top: API verification and metadata. Middle: LLM consistency checks on the cited paper's content. Bottom: bibliography verification of the cited paper's own references. When both consistency and metadata scores are available, they blend (60\% consistency, 40\% metadata). All signals are intrinsic; no popularity metrics. Clamped to $[0, 1]$.}
\label{tab:lightweight}
\end{table}

All signals are collected automatically during the default pipeline. API verification and metadata signals require no GPU. Consistency signals come from the same LLM checks that run on cited papers (step 4). Bibliography signals come from the network-only bibliography verification stage. Documents exceeding a configurable size threshold (default ${\sim}$50 pages) are excluded from consistency checks and receive neutral scores (1.0) so the threshold does not penalize citing books.

\subsection{Convergent Reliability}

Individual check signals each catch one failure mode. The strongest signal comes from convergence: when multiple independent checks flag the same reference, it is almost certainly unreliable.

The \texttt{reference-unreliable} check (Table~\ref{tab:checks}) aggregates metadata integrity, claim verification, and bibliography signals into a single reliability score per reference, firing a warning when the score drops below a threshold. This check and SciLint Score are parallel consumers of the same signals: the check surfaces actionable warnings; the score incorporates the same data as a number.

\subsection{Matching Engine}
\label{sec:matching-engine}

When no structured identifier is available, the pipeline must identify the correct paper from free-text metadata alone. This is especially important for AI-assisted writing, where LLMs routinely introduce metadata errors: wrong publication year, author name variations, truncated titles, and confused venues.

The engine queries each API with a deliberately loose search (first-author surname + title as free text) to retrieve up to 10 candidates. Structured API filters (year, author name) are intentionally avoided: if the bibliography entry contains a wrong year, a year filter would exclude the correct paper at the API level. All scoring happens client-side after retrieval, using the product of four graduated signals: title similarity, author overlap (with name variant expansion for initials, reversed order, and transliteration), quadratic year penalty (tolerating $\pm$1 year while crushing large mismatches), and venue tiebreaking. The best candidate above threshold (0.70) is accepted; below threshold, the reference is classified as T3 (not found).

\subsection{Integrity Checks}

\begin{table}[ht]
\centering
\small
\begin{tabularx}{\textwidth}{lX}
\toprule
\textbf{Check} & \textbf{What it catches} \\
\midrule
\multicolumn{2}{l}{\emph{Text checks (deterministic, no services)}} \\
\quad \texttt{dangling-cite} & \texttt{\textbackslash cite\{key\}} has no matching bib entry \\
\quad \texttt{dangling-ref} & \texttt{\textbackslash ref\{X\}} has no matching \texttt{\textbackslash label\{X\}} \\
\quad \texttt{unreferenced-figure} & Figure label defined but never cross-referenced \\
\addlinespace
\multicolumn{2}{l}{\emph{Pairwise LLM checks (section pairs, vLLM)}} \\
\quad \texttt{cross-section-consistency} & Numbers, claims, framing drift between sections \\
\quad \texttt{structure-promises} & Promised contributions not delivered \\
\addlinespace
\multicolumn{2}{l}{\emph{Full-paper LLM checks (shared APC prefix, vLLM)}} \\
\quad \texttt{numbers-vs-tables} & Numbers in text contradict the corresponding table \\
\quad \texttt{percentages-sum} & Reported percentages do not sum to 100\% \\
\quad \texttt{sample-size-consistency} & Sample size $N$ differs across sections without explanation \\
\quad \texttt{arithmetic-consistency} & Stated totals do not match their components \\
\quad \texttt{causal-language-audit} & Causal claims unsupported by study design \\
\quad \texttt{abstract-body-alignment} & Abstract makes factual claims the body contradicts \\
\quad \texttt{statistical-reporting} & Statistical results contradict their verbal interpretation \\
\addlinespace
\multicolumn{2}{l}{\emph{Figure checks (VL model + shared APC prefix, vLLM)}} \\
\quad \texttt{caption-vs-content} & Caption does not match the visual content of the figure \\
\quad \texttt{text-vs-figure} & Text describes a figure differently from what it shows \\
\quad \texttt{axis-label-consistency} & Axis labels or units mismatch the text \\
\quad \texttt{figure-data-vs-table} & Same data in a figure and table disagrees \\
\addlinespace
\multicolumn{2}{l}{\emph{Reference checks (APIs + vLLM)}} \\
\quad \texttt{reference-exists} & Not found in CrossRef, OpenAlex, S2, Open Library, or LoC \\
\quad \texttt{reference-accuracy} & Metadata mismatch against canonical records \\
\quad \texttt{retracted-cite} & Reference in Retraction Watch database \\
\quad \texttt{reference-unreliable} & Low aggregate reliability across convergent signals \\
\quad \texttt{claim-support} & Cited paper doesn't support the claim \\
\quad \texttt{cite-purpose} & Citation has no argumentative role (Table~\ref{tab:cite_functions}) \\
\addlinespace
\multicolumn{2}{l}{\emph{Prose check (vLLM)}} \\
\quad \texttt{prose-quality} & Grammar errors and wrong-word slips (e.g.\ ``comprise of'') \\
\bottomrule
\end{tabularx}
\caption{Twenty-three checks, all run by \texttt{sciwrite-lint check}. Three deterministic text checks require no services. The eleven full-paper checks (seven numerical + four figure) share a single APC-cached prefix containing the paper body and vision model figure descriptions. Figure checks are skipped deterministically when no figure descriptions are available. Figure checks use a two-model pipeline: Qwen3-VL (2B or 8B) describes, Qwen3 8B reasons. API checks require network. \texttt{prose-quality} reviews grammar and word choice; hedging, passive voice, and stylistic preference are deliberately not flagged.}
\label{tab:checks}
\end{table}


\section{SciLint Score}
\label{sec:scilint}

Citation count answers: ``how many people noticed this?'' SciLint Score answers: ``is the evidence real, and do the claims matter?'' The integrity component (the left factor below) is the core of the tool and is evaluated in this paper. The contribution component (the right factor) is an experimental extension released for community development (Section~\ref{sec:contribution}); when not explicitly invoked, contribution defaults to 1.0 and the score reduces to integrity alone.

For a paper $p$ with references $r_1, \ldots, r_n$:

\begin{equation}
S(p) = \underbrace{I(p)}_{\text{internal}} \;\times\; \underbrace{\operatorname{wmean}_{i} \bigl[ w_i \cdot V(p, r_i) \cdot R(r_i) \bigr]}_{\text{referencing quality}} \;\times\; \underbrace{\sum_{a \in \mathcal{A}} \beta_a \cdot C_a(p)}_{\text{contribution}}
\label{eq:scilint}
\end{equation}

where:
\begin{itemize}[nosep]
    \item $I(p)$: internal consistency (fraction of non-error findings in the manuscript),
    \item $V(p, r_i)$: claim verification score for reference $r_i$ (from SUPPORTS\,=\,1.0 to NOT\_SUPPORTED\,=\,0.0),
    \item $w_i$: citation purpose weight (Table~\ref{tab:cite_functions}; evidence $>$ example $>$ context),
    \item $\operatorname{wmean}$: weighted mean over references, with $w_i$ as weights,
    \item $R(r_i)$: reliability of the cited paper (Table~\ref{tab:lightweight}; blends metadata, consistency, and bibliography signals),
    \item $C_a(p)$: contribution score on axis $a$ (Section~\ref{sec:contribution}),
    \item $\beta_a$: contribution axis weights (equal by default; configurable per venue or discipline).
\end{itemize}

\begin{figure}[ht]
\centering
\begin{subfigure}[b]{0.32\textwidth}\centering\includegraphics[width=\textwidth]{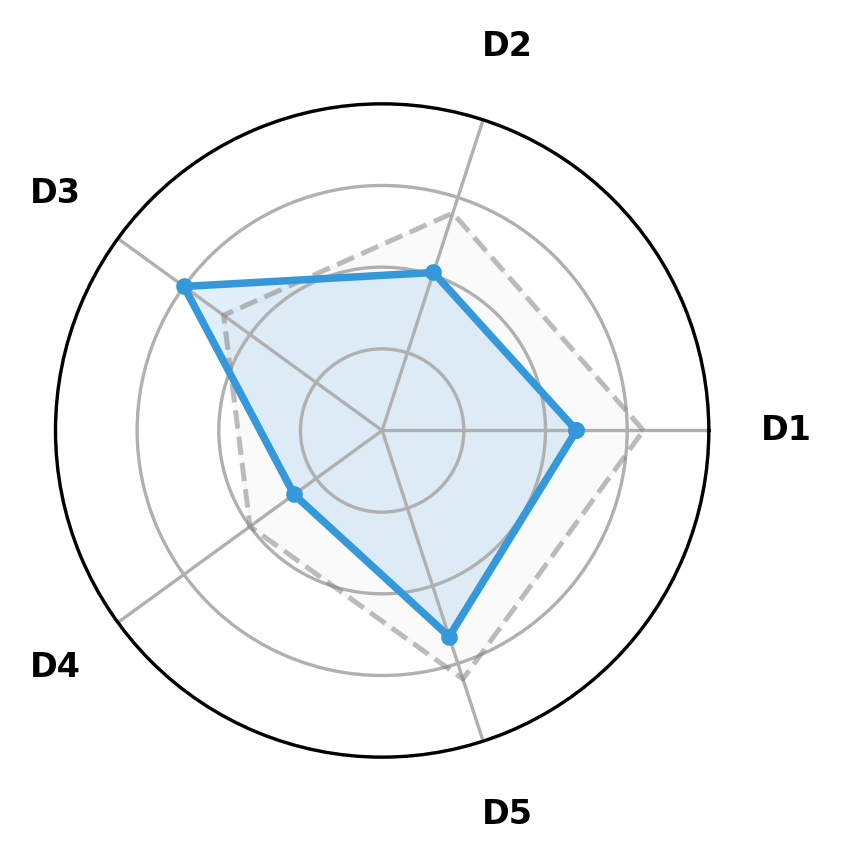}\caption{LIGO (2016): D1--D5 radar.}\end{subfigure}
\begin{subfigure}[b]{0.32\textwidth}\centering\includegraphics[width=\textwidth]{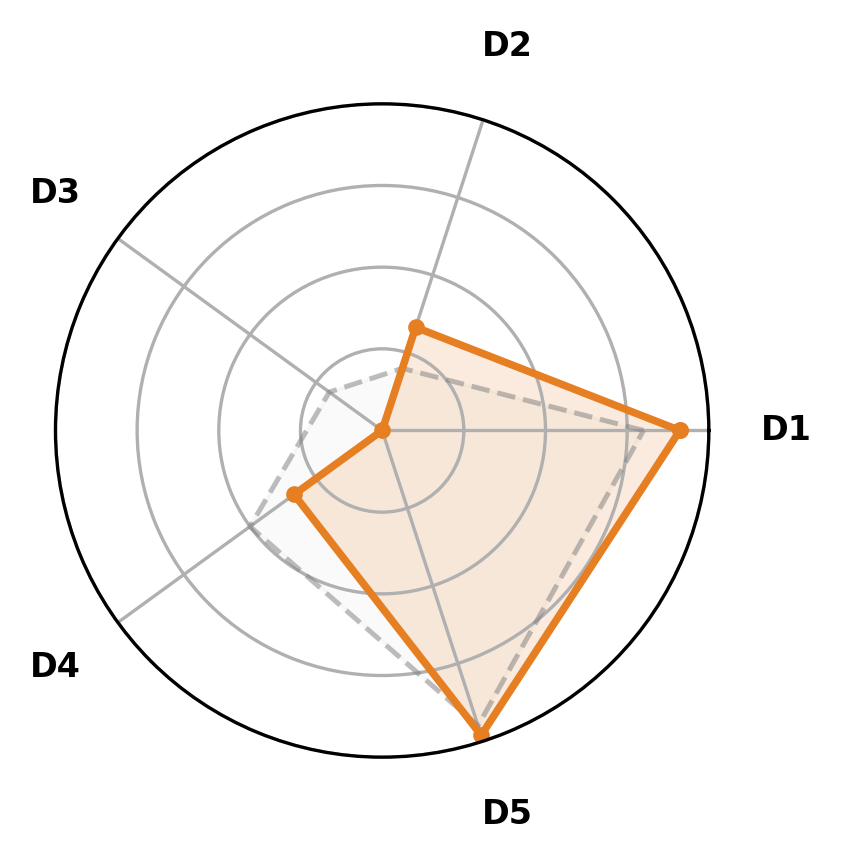}\caption{Camerer (2018): D1--D5 radar.}\end{subfigure}
\begin{subfigure}[b]{0.32\textwidth}\centering\includegraphics[width=\textwidth]{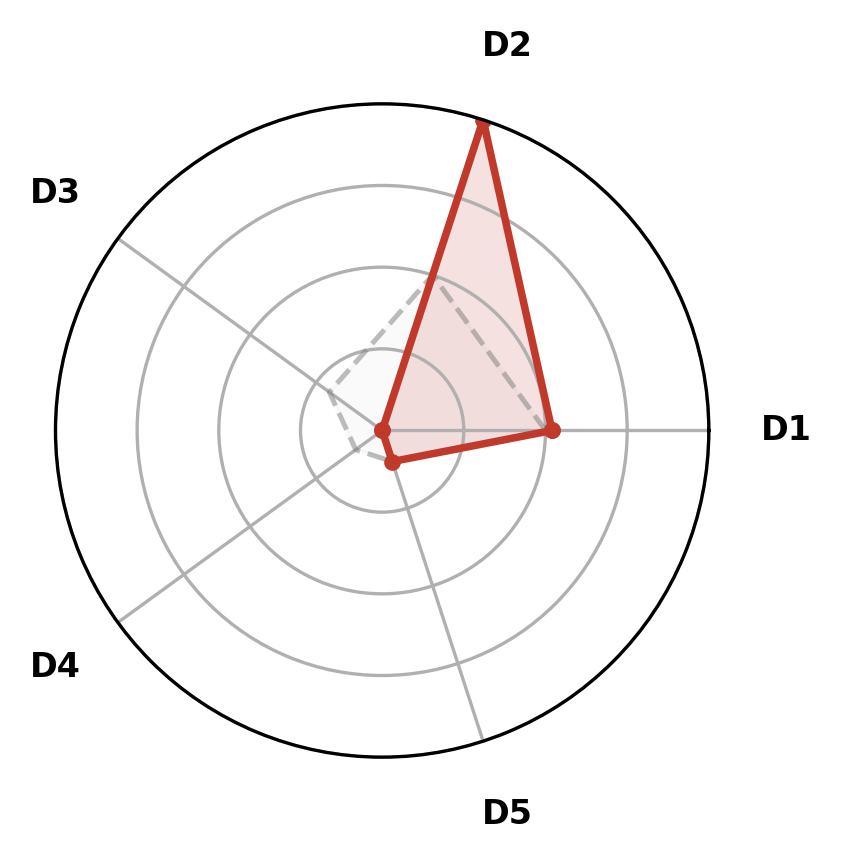}\caption{LaCour (2014): D1--D5 radar.}\end{subfigure}

\smallskip
\footnotesize D1 = Empirical content (Popper), D2 = Progressiveness (Lakatos), D3 = Unification (Kitcher), D4 = Problem-solving (Laudan), D5 = Test severity (Mayo). Dashed gray = target (human judgment); solid = SciLint Score after calibration.
\caption{Contribution profiles for 3 of the 20 calibration papers. Left: Nobel Prize discovery (balanced). Center: honest replication (high severity, low progressiveness, by design). Right: retracted fraud (suspicious progressiveness spike, near-zero severity).}
\label{fig:radar}
\end{figure}

The score is multiplicative: a paper must be both trustworthy \emph{and} substantial. Honestly-cited but vacuous work scores low on contribution. Brilliant arguments on fabricated evidence score low on integrity. Contribution defaults to 1.0 when not explicitly assessed, so the multiplicative structure engages only when \texttt{sciwrite-lint contributions} is run. The full profile (integrity and contribution subscores) can be displayed as a radar plot for diagnostics (Figure~\ref{fig:radar}).

An additional penalty applies when a paper makes bold progressive claims without methodological self-awareness: if problem-solving effectiveness is near zero ($< 0.1$) while progressiveness is high ($> 0.5$), the contribution score is multiplied by a dampening factor of 0.50--0.75, scaling with how bold the claims are.


\section{Beyond Integrity: Contribution Assessment (Experimental)}
\label{sec:contribution}

Integrity verification answers ``is the evidence real?'' but not ``does it matter?'' A paper can pass every check yet make no falsifiable predictions, solve no open problems, and test nothing severely. We sketch a direction for a contribution component and invite the community to develop it.

Five frameworks from philosophy of science can be operationalized as computable structural properties of arguments:

\paragraph{Empirical content (Popper).} Fraction of claims that are specific and falsifiable. ``Performance improves'' scores lower than ``F1 increases by 3--5\% on dataset X under condition Y.''

\paragraph{Progressiveness (Lakatos).} Ratio of novel predictions to ad-hoc accommodations.

\paragraph{Explanatory unification (Kitcher).} Citation-graph bridging across research communities. The verification pipeline provides the graph; clustering reveals whether a paper connects disconnected communities.

\paragraph{Problem-solving effectiveness (Laudan).} Problems claimed solved vs.\ limitations acknowledged. A paper claiming 5 solutions with zero limitations is suspicious.

\paragraph{Test severity (Mayo).} Ablations, strong baselines, alternative explanations addressed.

A deliberate exclusion: \emph{novelty}. Novelty is unreliable for cross-sectional assessment because it can only be judged in retrospect. Novoselov and Geim's first graphene paper \citep{novoselov2004} was a curiosity in 2004 and a Nobel Prize in 2010. We replace novelty with Lakatos's \emph{progressiveness}, which is measurable at the time of publication.

Each claim is classified along five dimensions (type, specificity, testability, support, scope) using the same local-LLM infrastructure. The axis weights $\beta_a$ in Equation~\ref{eq:scilint} should depend on paper type: a review paper should weight Kitcher (unification) heavily and Mayo (test severity) near zero; an experimental paper is the opposite.

To our knowledge, no prior work operationalizes these five frameworks as computable paper properties. The closest work (altmetrics, Semantic Scholar influence scores, scite.ai \citep{nicholson2021scite}) measures attention or agreement, not structural quality of arguments.


\section{Implementation: sciwrite-lint}
\label{sec:design}

The pipeline and scoring framework described above are packaged as sciwrite-lint, an open-source tool (\texttt{pip install sciwrite-lint}).

\paragraph{Structured pipeline, not agentic system.} The LLM is a semantic processing engine, not a knowledge source or autonomous agent. Reference existence is checked via deterministic API lookups. Metadata accuracy uses fuzzy string matching with fixed thresholds. Retraction status is a database lookup. The LLM handles only tasks requiring language understanding: given a retrieved passage and a claim, does the passage support the claim? Each LLM call receives structured input and produces structured output (JSON classification). The LLM never decides what to search, which tool to call, or how to plan. This is a deliberate alternative to agentic architectures: structured flows are reproducible, auditable, and run on a single consumer GPU.

\paragraph{Local by design.} Deterministic database lookups require no GPU. Local-LLM checks use Qwen3 8B (FP8 quantized) via vLLM on a consumer GPU with 16\,GB+ VRAM and FP8 hardware support (NVIDIA Ada Lovelace or Hopper; e.g.\ RTX 4070 Ti / 4080 / 4090, RTX 4000 / 5000 / 6000 Ada, H100). The pipeline was developed on a single workstation (Windows/WSL2, 64\,GB RAM, NVIDIA RTX 4000 Ada 20\,GB VRAM) and should also run on native Linux. No manuscripts leave the researcher's machine. A centralized verification authority has the same structural problem as journal gatekeeping: whoever controls the tool controls the verdict. Local computation eliminates this.

\paragraph{Network security model.} While manuscripts stay local, the verification pipeline contacts roughly 16 external APIs with citation metadata (DOIs, titles, author names). No paper content is sent externally. All external communication uses HTTPS with rate limiting. Internally, the pipeline defends against malicious content: \texttt{defusedxml} blocks entity-expansion attacks, downloads enforce size limits, URL redirects are validated against DNS-resolved IP addresses. LLM output is parsed as structured data and never fed back into network requests, file paths, or shell commands; the pipeline completes all external API calls before any LLM invocation. The LLM is a terminal node: it consumes retrieved data and produces display output, with no path back to the network. Conversely, cited papers parsed by GROBID are untrusted input to the LLM: a malicious PDF could embed prompt injection payloads in its text. The pipeline mitigates this with XML delimiters separating trusted instructions from document content, anti-injection directives in system prompts, and strict JSON schema enforcement on all LLM output.

\paragraph{The Verification Prior Hypothesis.} For verification of scientific manuscripts, smaller language models may be a better fit than larger ones. Novel findings may contradict established knowledge; a model with strong priors is exactly wrong when verifying a paper that challenges the consensus. Three lines of evidence support this. First, inverse scaling causes performance to worsen with scale on certain tasks \citep{mckenzie2023inversescaling,wei2023ushaped}, and sycophancy increases with model size \citep{sharma2024sycophancy}. Second, small specialized models can match or outperform GPT-4 on fact-checking: MiniCheck (770M) reaches GPT-4-level accuracy at a fraction of the cost \citep{tang2024minicheck}, building on earlier small-model fact-checkers such as AlignScore (355M) and fine-tuned DeBERTa that established strong claim-verification performance at sub-billion-parameter scale \citep{zha2023alignscore,kosprdic2024claimverification}. Third, intrinsic self-correction typically degrades output because the model shares the same blind spots as the generator and the evaluator \citep{huang2024selfcorrect,kamoi2024selfcorrection}.

\paragraph{Maximizing a single GPU.} The pipeline is engineered to saturate a consumer GPU rather than require multiple. Three models share one GPU through \textbf{sequential scheduling}: (1) the vision-language model (Qwen3-VL-2B at ${\sim}$4\,GB FP16, or 8B at ${\sim}$8\,GB FP8, user-selectable) loads, describes all figures, and unloads; (2) the embedding model (Snowflake Arctic Embed M v2.0, ${\sim}$1.2\,GB) loads during parsing and unloads; (3) vLLM serves Qwen3 8B for all text-based checks with FP8 KV cache (${\sim}$131K tokens on 20\,GB VRAM), \textbf{automatic prefix caching} (APC), and \textbf{chunked prefill}. When the vision backend is vLLM as well, the text container is stopped briefly so the vision container can claim the GPU, then restarted; the same container-swap path works on WSL2 and native Linux. The embedder uses PyTorch's memory-efficient scaled dot-product attention (SDPA, O$(N)$ memory rather than O$(N^2)$) and bf16 weights (cosine similarity $\geq 0.9998$ vs fp32), pre-warmed during GROBID parse; this is roughly $7\times$ faster than the default fp32 path on cache-cold runs. Pipeline stages are scheduled to avoid KV cache contention: consistency checks on cited papers run first (thinking=low, ${\sim}$5--10\,s), then claim verification (thinking disabled) runs concurrently with bibliography verification (network-only). The cite-purpose classifier exploits APC most aggressively: all 30--60 queries share a ${\sim}$10--20K token prefix, computed once and reused via hash lookup. \textbf{Admission control is dynamic}: a controller polls vLLM's \texttt{/metrics} endpoint and resizes the in-flight cap based on observed KV-cache utilization and queue depth, instead of a static client-side semaphore. Wire-side JSON schema length constraints (\texttt{maxLength}, \texttt{maxItems}) are stripped before submission so xgrammar's constrained-decoding fast path stays engaged; a single field-level \texttt{maxLength} collapsed throughput by roughly $3\times$ at concurrency 60 in our benchmarking. Pydantic still validates the decoded JSON post-response.

\paragraph{Eval-driven prompt engineering.} Every prompt is optimized empirically: dedicated accuracy evals with thinking-preset sweeps, not intuition. The key principle: the optimal thinking budget varies by task type and \emph{cannot be assumed from perceived task complexity}. Each of the seven vLLM prompts was swept independently; the results split cleanly. Classification and extraction tasks (pattern matching over short input) perform best with thinking disabled. The model overthinks, rationalizing its way into wrong categories. Multi-dimension reasoning (classifying a claim along five philosophical axes simultaneously) benefits from moderate thinking. Contradiction detection needs minimal thinking. Prompt wording follows one rule: sharper category descriptions beat additional instructions or few-shot examples. When the model confuses two categories, the fix is making the discriminating feature explicit in the description (``explains WHAT a concept means'' vs.\ ``names WHO created it''), not adding more text. Ten eval systems form a hierarchy from unit tests (700+, no services) through synthetic detection (P/R/F1 per check), calibration (20 papers, 38 ordinal constraints), to real-world corpus evaluation (30 unseen papers). Every number reported in this paper is reproducible by running the corresponding eval command; the evals are released as part of the tool.

\paragraph{Open-weights reproducibility.} Pinning to an open-weights model version is deliberate. A verification tool whose results change because the provider updated their API is not trustworthy. The framework is model-agnostic: the community can swap, fine-tune, or distill models. SemanticCite \citep{haan2025semanticcite} demonstrates that QLoRA fine-tuning of Qwen3 achieves strong results at 4B parameters.

\paragraph{Robust LLM input.} Manuscript text is preprocessed with pandoc before any LLM check sees it: citations collapse to a \texttt{[CITE]} marker, cross-references to \texttt{[REF]}, math renders as Unicode, and stranded punctuation from elided LaTeX is cleaned up. This isolates the LLM from idiosyncratic LaTeX surface form and removes a common source of nuisance findings.

\paragraph{Composable Python API.} The open-access acquisition layer is exposed as a public Python API (\texttt{down\-load\_pdf}, \texttt{fetch\_web}, \texttt{search\_\-by\_\-title}) so that reference managers, RAG systems, and writing agents can reuse the validation chain without invoking the linter. Typed exceptions (\texttt{Sci\-Write\-Lint\-Error}, \texttt{LLM\-Connection\-Error}) and standard 0/1/2 exit codes make the tool easy to script around.

\paragraph{Configurable rules and transparent diagnostics.} Every check can be enabled, disabled, or reconfigured via \texttt{.sciwrite-lint.toml}; \texttt{sciwrite-lint check --checks ID[,ID...]} runs an explicit subset for fast iteration. Two audiences: humans read terminal output; AI writing agents consume JSON and run the linter in a write$\to$check$\to$fix$\to$recheck loop. Every finding carries a \emph{message} (what was found), a \emph{context} (why), and a \emph{level} that distinguishes manuscript issues from tool limitations. Operational state (LLM unavailability, an incomplete vision pass, a failed parse, an unexpected exception inside a check) is routed to a separate \texttt{system\_issues} bucket with purpose-built rule IDs (\texttt{llm-unavailable}, \texttt{vision-incomplete}, \texttt{parse-failed}, \texttt{internal-error}) and never inflates manuscript-finding totals or the SciLint Score: the score reflects manuscript quality only, regardless of how the linter's run went.

\paragraph{Structured identifiers and metadata resilience.} When bibliography entries include DOIs, arXiv IDs, PMIDs, ISBNs, or LCCNs, verification is deterministic: one API call confirms existence and returns canonical metadata. Structured identifiers enable batching: DOIs via OpenAlex (up to 200 per call), arXiv IDs and PMIDs via Semantic Scholar (500 per call), ISBNs via Open Library, LCCNs via Library of Congress. Open abstract coverage is unstable across publishers, but core verification depends on metadata (titles, authors, DOIs), not abstracts.


\section{Evaluation}

SciLint Score is a framework with a first implementation, not a claim of optimality. The weights, classification prompts, and scoring thresholds are initial choices that future work will improve. We report the current implementation's accuracy honestly, including known model limits.

\subsection{Framework vs.\ Optimized Implementation}

The current implementation is deliberately minimal: a single general-purpose model (Qwen3 8B) with prompt engineering only, no fine-tuning, no per-paper-type weight calibration. Equal contribution weights ($\beta_a = 0.2$) are used regardless of paper type. The contribution is the framework (the operationalization of philosophy of science into computable properties), not any particular configuration. Each component is independently improvable by the community.

\begin{table}[ht]
\centering
\small
\begin{tabularx}{\textwidth}{llX}
\toprule
\textbf{Choice} & \textbf{Value} & \textbf{What it controls} \\
\midrule
$\beta_a$ & 0.2 each (equal) & Contribution axis weights; not yet calibrated per paper type \\
$w_i$ & 0.2--1.0 (Table~\ref{tab:cite_functions}) & Citation purpose weights in referencing quality \\
T1/T2/T3 scores & 0.9/0.7/0.3 & Reference integrity by verification tier \\
Mismatch penalty & $-0.1$ each & Per-field metadata accuracy deduction \\
Retracted score & 0.0 & Hard floor for retracted references \\
Base model & Qwen3 8B & All LLM calls; no fine-tuning applied \\
Embedding model & Snowflake Arctic Embed M v2.0 & Claim retrieval; shares VRAM with vLLM \\
Title match threshold & 0.80 & Fuzzy similarity below this $\to$ ERROR \\
Calibration method & Prompt iteration only & No training data, no model selection search \\
\bottomrule
\end{tabularx}
\caption{Key design choices in the current implementation. Each is a starting point for community optimization.}
\label{tab:choices}
\end{table}

\subsection{Calibration Against Known Papers}
\label{sec:calibration}

We selected 20 open-access papers ($\leq$25 pages) spanning seven domains (CS/ML, physics, biology, social science, medicine, ecology, economics) and defined 38 ordinal constraints (pairwise ranking expectations such as ``LIGO (2016) should score higher than LK-99 (2023)''). Papers range from Nobel Prize work (Novoselov 2004, Abbott et al.\ 2016, Jinek et al.\ 2013, Jumper et al.\ 2021) to retracted fraud (Shoukat et al.\ 2024, LaCour \& Green 2014). The expected rankings are our own assessment, not an independent benchmark, and reasonable people may disagree. The value is in the methodology: making expectations explicit and checking them systematically.

\begin{table}[ht]
\centering
\small
\begin{tabular}{lrrrrrrr}
\toprule
\textbf{Paper} & \textbf{Score} & \textbf{Int.} & \textbf{Emp.} & \textbf{Prog.} & \textbf{Unif.} & \textbf{Prob.} & \textbf{Sev.} \\
\midrule
LIGO (2016) & 0.570 & 1.00 & 0.59 & 0.51 & 0.75 & 0.33 & 0.66 \\
Reinhart-Rogoff (2010) & 0.553 & 1.00 & 0.77 & 0.33 & 0.67 & 0.33 & 0.67 \\
Graphene (2004) & 0.536 & 1.00 & 0.73 & 0.80 & 0.00 & 0.25 & 0.90 \\
Transformer (2017) & 0.512 & 1.00 & 0.54 & 0.36 & 0.86 & 0.25 & 0.55 \\
Camerer (2018) & 0.512 & 1.00 & 0.91 & 0.33 & 0.00 & 0.33 & 0.98 \\
\midrule
LK-99 (2023) & 0.281 & 1.00 & 0.71 & 0.81 & 0.00 & 0.00 & 0.83 \\
RECOVERY (2020) & 0.278 & 1.00 & 0.56 & 0.27 & 0.00 & 0.25 & 0.31 \\
Wu survey (2021) & 0.173 & 1.00 & 0.40 & 0.33 & 0.00 & 0.00 & 0.13 \\
LaCour (2014) & 0.162 & 1.00 & 0.52 & 1.00 & 0.00 & 0.00 & 0.10 \\
\bottomrule
\end{tabular}
\caption{SciLint Score calibration: top 5 and bottom 4 of 20 papers (standalone mode; full results via \texttt{eval-calibration}). Standalone mode sets integrity to 1.0 to isolate contribution scoring from API variability.}
\label{tab:calibration}
\end{table}

30 of 38 ordinal constraints pass (79\%). Nobel Prize papers rank in the top quartile; retracted papers rank in the bottom quartile. The bold-claims penalty correctly identifies LK-99: despite high empirical content (0.71) and progressiveness (0.81), its zero problem-solving score triggers a penalty. Honest replication studies (Camerer 2018) are not penalized for low progressiveness; they compensate with high test severity (0.98).

The unification axis returns zero for 8 of 20 papers due to an algorithmic limitation (Section~\ref{sec:limitations}). Clinical trial methodology (RECOVERY 2020) is underscored on test severity despite pre-registration and large sample size, suggesting the taxonomy prompt does not yet adequately recognize RCT-specific evidence patterns.

Calibration as methodology: the calibration set, ordinal constraints, and iteration history are released as part of the tool (\texttt{eval-calibration}). Users can add domain-specific papers and constraints. The key discipline: every prompt or weight change must be a general improvement, never a special case targeting a specific calibration paper.

\subsection{Real-World Corpus Evaluation}
\label{sec:realworld}

Calibration validates \emph{scoring}; this section validates \emph{robustness}. We downloaded 30 papers (13~arXiv, 17~bioRxiv) spanning computer science, physics, biology, neuroscience, ecology, and medicine, with no paper examined before inclusion.

\paragraph{Injection recall.} We injected 68 synthetic errors (fake \verb|\cite{}| commands and broken cross-references). The linter detected 67 of 68: \texttt{dangling-cite} 38/38 (100\%) and \texttt{dangling-ref} 29/30 (96.7\%), for 98.5\% aggregate recall with zero false positives on both LaTeX and PDF input.

\paragraph{Full pipeline.} 27 of 30 papers completed successfully. SciLint Scores ranged from 0.071 to 0.533 (mean 0.254). The pipeline produced 1,755 findings across 27 papers (65 per paper), dominated by \texttt{reference-exists} (1,334, most from PDF papers whose references lack DOIs and depend on title search) and \texttt{reference-accuracy} (258).

\paragraph{False positive analysis.} We adjudicated the top 20 findings per paper using Claude Sonnet as an independent judge (379 findings total): 14\% TP, 60\% FP, 26\% UNCERTAIN. This FPR was measured before the matching engine was implemented. Deterministic text checks have near-zero FPR; the high overall FPR is concentrated in reference-DB checks on PDF papers where references lack DOIs and the pipeline relied on naive title search. Two responses now address this: (1)~the pipeline extracts structured identifiers from GROBID's TEI XML; when a DOI is available, verification is deterministic with zero FPR; (2)~the matching engine (Section~\ref{sec:matching-engine}) replaces naive title matching with composite scoring. A dedicated evaluation (\texttt{eval-real-world matching}) tests 17 degradation scenarios per paper.

\paragraph{Complementary methodologies.} Calibration found that the unification axis was zero on a large fraction of papers and that Laudan defaulted poorly for missing sections. Real-world eval found that API lookups silently failed and naive title search produced wrong matches, motivating the matching engine. Both are released as part of the tool.


\section{Related Work}
\label{sec:related}

\paragraph{Research assessment.} Beyond the attention metrics discussed in Section~\ref{sec:intro}, PageRank \citep{pagerank1999} and its citation-network application \citep{chen2007pagerankgems}, the h-index \citep{hirsch2005hindex}, and altmetrics belong to the same family: they measure influence, not properties of the paper itself. scite.ai \citep{nicholson2021scite} classifies 1.5B citation statements as supporting/contrasting; this is closer to what we want, but still measures what others say about a paper. SciScore \citep{menke2020sciscore} automates methods-section scoring for rigor and transparency; \citet{ripeta2021} propose RipetaScore for research practices, professionalism, and reproducibility. These check reporting compliance, not evidence or arguments. Automated novelty detection \citep{wang2025novelty} identifies new concepts, but novelty is unreliable for cross-sectional assessment (Section~\ref{sec:contribution}).

\paragraph{Citation verification.} Existing tools check the manuscript's own references; none follow citations into referenced papers (Table~\ref{tab:tools}, in the Introduction).

CiteAudit \citep{yuan2026citeaudit} proposes a multi-agent verification pipeline for citation auditing. At the time of writing, the announced benchmark and framework release are not yet publicly available. CiteVerifier \citep{xu2026ghostcite} is open-source (existence checks only, primarily CS coverage). RefChecker \citep{russinovich2025refchecker} is \texttt{pip}-installable (existence checks via CrossRef, OpenAlex, and Semantic Scholar, plus LLM web search for flagged entries). SemanticCite \citep{haan2025semanticcite} verifies claims with fine-tuned Qwen3 (1.7B/4B QLoRA) via a five-stage inference pipeline (chunking, claim extraction, hybrid retrieval, reranking, classification), but operates on individual citation--reference pairs and provides no manuscript-level ingestion. None of these four tools support retraction detection, bibliography verification, or citation purpose assessment.

\paragraph{Publisher screening and automated peer review.} Publisher tools (Problematic Paper Screener \citep{cabanac2021tortured}, STM Integrity Hub, Springer Nature) ask ``was this AI-generated?'' (a detection problem). SciLint Score asks ``is the evidence correct?'' (a verification problem). Detection and verification are orthogonal. LLM-based review systems that generate full reviews \citep{beel2025sakana} aim to replicate subjective quality judgments. SciLint Score measures specific structural properties that are verifiable and reproducible.

\paragraph{Philosophy of science as computation.} Coherence-based models such as Thagard's ECHO \citep{thagard1989explanatory} evaluate \emph{theory competition}, not individual paper properties. Bayesian confirmation approaches require prior probabilities that do not exist for novel findings. The Stanford POPPER framework \citep{popper2025stanford} operationalizes Popper for running experiments, not assessing papers. The F-index \citep{fidx2024} proposes falsifiability as a metric but requires author-provided statements and is not implemented. No prior work automates Lakatos, Kitcher, Laudan, or Mayo as computable paper properties.

\paragraph{Claim verification and domain-specific checking.} SciFact \citep{wadden2020scifact}: 1,409 expert-annotated claims. MiniCheck \citep{tang2024minicheck}: GPT-4-level accuracy at 400$\times$ lower cost. statcheck \citep{nuijten2016statcheck} found inconsistent $p$-values in approximately half of 250,000+ psychology papers, a precedent that SciLint Score extends from statistical reporting to the full evidence chain.

\paragraph{Linting.} Lint was introduced at Bell Labs in 1978 as a static checker for C source \citep{johnson1978lint}. Modern linters (ESLint, Ruff) are integral to software development. scicode-lint \citep{samsonau2025scicodelint} applies the paradigm to ML methodology bugs. sciwrite-lint extends it from single-file checking to automatic bibliography verification with accountability scoring.


\section{Limitations and Future Work}
\label{sec:limitations}

\paragraph{Speed.} An initial run on a paper with 50 references takes up to 30 minutes, dominated by claim verification; subsequent cached runs complete in minutes. Every component uses open-weights models that can be swapped, fine-tuned, or distilled as faster alternatives become available.

\paragraph{Figure analysis.} The vision pipeline extracts raster images from PDFs and renders TikZ/pgfplots figures from compiled LaTeX. Vector graphics in PDF-only input (no source \texttt{.tex}) are not covered; they are drawing commands, not embedded images. The 8B VL model detects 100\% of injected figure errors (62 cases); the 2B model detects 85\%. The downstream figure checks that reason over these descriptions have not yet been evaluated at scale; false positive rates for caption-vs-content and text-vs-figure need calibration.

\paragraph{Preliminary evaluation.} The pilot calibration (20 papers, 38 constraints) and real-world corpus (30 papers) validate scoring and robustness respectively, but key questions remain: optimal weights across disciplines, score distribution across venues, and correlation with retraction status. A larger calibration set with broader domain coverage would strengthen the methodology.

\paragraph{Contribution assessment.} The five-axis framework is experimental. Qwen3 8B reliably classifies claim type, support, and scope, but struggles with testability and specificity (e.g., classifying logical necessities as falsifiable). The philosophical foundations themselves have known critiques: Popper's falsifiability (Duhem-Quine), Lakatos's progressive/degenerative distinction, Kitcher's domain-dependent unification. We use these as operationalizable heuristics, not definitive measures.


\section{Applications}

\paragraph{Linting AI-assisted manuscripts.} The primary use case: run \texttt{sciwrite-lint check} before submission. The tool catches what AI-assisted writing systematically introduces and human review systematically misses: fabricated references, unsupported claims, cross-section contradictions. The internal-consistency checks (Table~\ref{tab:checks}) target the same failure mode \citet{laban2026delegate} document at scale: when an LLM is delegated a long edit, errors accumulate silently across the document and are invisible in the local diff. Unlike detection (``did an AI write this?''), linting asks ``is this correct?''

\paragraph{Pre-reviewer for journals, only reviewer for preprints.} Editors see the SciLint Score before assigning reviewers; reviewers see what has been verified and focus effort on what requires judgment. On arXiv and bioRxiv, where there is no peer review, SciLint Score is the only quality signal between the author and the public record.

\paragraph{Feedback for writers.} The radar plot provides what a good advisor provides: not ``needs more work'' but ``your claims are vague (Popper 0.3), you have no ablations (Mayo 0.2), your references are all from one community (Kitcher 0.1).'' The five contribution axes are themselves a scientific writing curriculum.

\paragraph{Ingestion filter for AI agents.} RAG systems and automated literature reviewers check SciLint Score before propagating claims. A paper with low integrity is not ingested. This closes the loop: when AI reads papers, the score filters what enters the knowledge base.


\section{Conclusion}

Citation verification at scale is structurally beyond what human review can deliver, and AI-assisted writing makes the gap more urgent.

sciwrite-lint applies the linting paradigm from software engineering to scientific manuscripts. The pipeline verifies that references exist, checks claim support against the cited paper's content, and traces evidence one level deeper into cited papers' bibliographies. The same linting workflow extends to internal consistency: numbers vs.\ tables, abstract vs.\ body, figure vs.\ text. It runs locally on a single consumer GPU, fast enough to re-lint between revisions, and serves both authors during drafting and reviewers as an automated first pass.

We propose SciLint Score to go further: combining integrity verification with contribution assessment that operationalizes five frameworks from philosophy of science. The integrity component is evaluated in this paper; the contribution component is released as experimental code for community development.

The framework is open-source because it must be. An accountability tool that cannot itself be audited fails its own standard.

\vspace{1em}
\noindent\textbf{Note on references.} Every reference in this paper is available in full text without institutional access or payment. While other influential work exists on relevant topics, we intentionally include only sources that any reader can obtain and read.

\bibliographystyle{plainnat}
\bibliography{references_v11}

\begin{thebibliography}{44}
\providecommand{\natexlab}[1]{#1}
\providecommand{\url}[1]{\texttt{#1}}
\expandafter\ifx\csname urlstyle\endcsname\relax
  \providecommand{\doi}[1]{doi: #1}\else
  \providecommand{\doi}{doi: \begingroup \urlstyle{rm}\Url}\fi

\bibitem[{AI Incident Database}(2023)]{samsung2023chatgpt}
{AI Incident Database}.
\newblock Incident 768: {Samsung ChatGPT} data leak.
\newblock \url{https://incidentdatabase.ai/cite/768/}, 2023.
\newblock Responsible AI Collaborative incident catalog. Reported leak of
  source code and internal meeting notes via ChatGPT, March 2023.

\bibitem[{American Society for Cell Biology}(2012)]{dora2012}
{American Society for Cell Biology}.
\newblock San {Francisco} declaration on research assessment ({DORA}).
\newblock \url{https://sfdora.org}, 2012.

\bibitem[Ansari(2026)]{ansari2026compound}
Jawad Ansari.
\newblock Compound deception in elite peer review.
\newblock \emph{arXiv preprint arXiv:2602.05930}, 2026.

\bibitem[Beel et~al.(2025)Beel, Vente, and Mahlich]{beel2025sakana}
Joeran Beel, Tobias Vente, and Christin Mahlich.
\newblock Evaluating sakana's {AI} scientist: Bold claims, mixed results.
\newblock \emph{arXiv preprint arXiv:2502.14297}, 2025.

\bibitem[Bienz et~al.(2026)Bienz, Pearson, and Garcia~de
  Gonzalo]{bienz2026mysterious}
Timothy Bienz, Arianna Pearson, and Sylvie Garcia~de Gonzalo.
\newblock The case of the mysterious citations.
\newblock \emph{arXiv preprint arXiv:2602.05867}, 2026.

\bibitem[Brembs et~al.(2013)Brembs, Button, and Munaf{\`o}]{brembs2013deep}
Bj{\"o}rn Brembs, Katherine Button, and Marcus Munaf{\`o}.
\newblock Deep impact: Unintended consequences of journal rank.
\newblock \emph{Frontiers in Human Neuroscience}, 7:\penalty0 291, 2013.
\newblock \doi{10.3389/fnhum.2013.00291}.

\bibitem[Cabanac et~al.(2021)Cabanac, Labb{\'e}, and
  Magazinov]{cabanac2021tortured}
Guillaume Cabanac, Cyril Labb{\'e}, and Alexander Magazinov.
\newblock Tortured phrases: A dubious writing style emerging in science.
  evidence of critical issues in engineered papers.
\newblock \emph{arXiv preprint arXiv:2107.06751}, 2021.

\bibitem[Chen et~al.(2007)Chen, Xie, Maslov, and Redner]{chen2007pagerankgems}
P.~Chen, H.~Xie, S.~Maslov, and S.~Redner.
\newblock Finding scientific gems with {Google}'s {PageRank} algorithm.
\newblock \emph{Journal of Informetrics}, 1\penalty0 (1):\penalty0 8--15, 2007.
\newblock \doi{10.1016/j.joi.2006.06.001}.

\bibitem[Dai et~al.(2021)Dai, Chen, Wan, Liu, Gong, and
  Wang]{dai2021references}
Can Dai, Quan Chen, Tao Wan, Fan Liu, Yanbing Gong, and Qingfeng Wang.
\newblock Literary runaway: Increasingly more references cited per academic
  research article from 1980 to 2019.
\newblock \emph{PLOS ONE}, 16\penalty0 (8):\penalty0 e0255849, 2021.
\newblock \doi{10.1371/journal.pone.0255849}.

\bibitem[Edwards and Roy(2017)]{edwards2017perverse}
Marc~A. Edwards and Siddhartha Roy.
\newblock Academic research in the 21st century: Maintaining scientific
  integrity in a climate of perverse incentives and hypercompetition.
\newblock \emph{Environmental Engineering Science}, 34\penalty0 (1):\penalty0
  51--61, 2017.
\newblock \doi{10.1089/ees.2016.0223}.

\bibitem[Haan(2025)]{haan2025semanticcite}
Sebastian Haan.
\newblock {SemanticCite}: Citation verification with {AI}-powered full-text
  analysis and evidence-based reasoning.
\newblock \emph{arXiv preprint arXiv:2511.16198}, 2025.

\bibitem[Hirsch(2005)]{hirsch2005hindex}
J.~E. Hirsch.
\newblock An index to quantify an individual's scientific research output.
\newblock \emph{Proceedings of the National Academy of Sciences}, 102\penalty0
  (46):\penalty0 16569--16572, 2005.
\newblock \doi{10.1073/pnas.0507655102}.

\bibitem[Huang et~al.(2024)]{huang2024selfcorrect}
Jie Huang et~al.
\newblock Large language models cannot self-correct reasoning yet.
\newblock In \emph{Proceedings of the Twelfth International Conference on
  Learning Representations (ICLR)}, 2024.

\bibitem[Huang et~al.(2025)]{popper2025stanford}
Yuxuan Huang et~al.
\newblock {POPPER}: An agentic framework for automated hypothesis validation
  via {Karl Popper}'s falsification.
\newblock \emph{arXiv preprint arXiv:2502.09858}, 2025.
\newblock ICML 2025.

\bibitem[Johnson(1978)]{johnson1978lint}
Stephen~C. Johnson.
\newblock Lint, a {C} program checker.
\newblock Technical Report~65, Bell Laboratories, 1978.
\newblock URL
  \url{https://wolfram.schneider.org/bsd/7thEdManVol2/lint/lint.pdf}.

\bibitem[Kamoi et~al.(2024)]{kamoi2024selfcorrection}
Ryo Kamoi et~al.
\newblock When can {LLMs} actually correct their own mistakes? {A} critical
  survey of self-correction of {LLMs}.
\newblock \emph{Transactions of the Association for Computational Linguistics},
  12:\penalty0 1417--1440, 2024.
\newblock \doi{10.1162/tacl_a_00713}.
\newblock arXiv:2406.01297.

\bibitem[Kosprdic et~al.(2024)]{kosprdic2024claimverification}
Milo{\v{s}} Kosprdic et~al.
\newblock Scientific claim verification with fine-tuned {NLI} models.
\newblock In \emph{Proceedings of the 16th International Conference on
  Knowledge Discovery and Information Retrieval (KDIR)}, pages 129--136, 2024.
\newblock \doi{10.5220/0012900000003838}.

\bibitem[Laban et~al.(2026)Laban, Schnabel, and Neville]{laban2026delegate}
Philippe Laban, Tobias Schnabel, and Jennifer Neville.
\newblock {LLMs} corrupt your documents when you delegate.
\newblock \emph{arXiv preprint arXiv:2604.15597}, 2026.

\bibitem[Larivi{\`e}re and Gingras(2010)]{lariviere2010matthew}
Vincent Larivi{\`e}re and Yves Gingras.
\newblock The impact factor's {Matthew} effect: A natural experiment in
  bibliometrics.
\newblock \emph{Journal of the American Society for Information Science and
  Technology}, 61\penalty0 (2):\penalty0 424--427, 2010.
\newblock \doi{10.1002/asi.21232}.

\bibitem[McKenzie et~al.(2023)]{mckenzie2023inversescaling}
Ian~R. McKenzie et~al.
\newblock Inverse scaling: When bigger isn't better.
\newblock \emph{Transactions on Machine Learning Research}, 2023.

\bibitem[Menke et~al.(2020)Menke, Roelandse, Ozyurt, Martone, and
  Bandrowski]{menke2020sciscore}
Joe Menke, Martijn Roelandse, Burak Ozyurt, Maryann Martone, and Anita
  Bandrowski.
\newblock The rigor and transparency index quality metric for assessing
  biological and medical science methods.
\newblock \emph{iScience}, 23\penalty0 (11):\penalty0 101698, 2020.
\newblock \doi{10.1016/j.isci.2020.101698}.

\bibitem[Mogull(2017)]{mogull2017quotation}
Scott~A. Mogull.
\newblock Accuracy of cited ``facts'' in medical research articles: {A} review
  of study methodology and recalculation of quotation error rate.
\newblock \emph{PLOS ONE}, 12\penalty0 (9):\penalty0 e0184727, 2017.
\newblock \doi{10.1371/journal.pone.0184727}.

\bibitem[Nasr et~al.(2023)Nasr, Carlini, Hayase, Jagielski, Cooper, Ippolito,
  Choquette-Choo, Wallace, Tram\`{e}r, and Lee]{nasr2023extracting}
Milad Nasr, Nicholas Carlini, Jonathan Hayase, Matthew Jagielski, A.~Feder
  Cooper, Daphne Ippolito, Christopher~A. Choquette-Choo, Eric Wallace, Florian
  Tram\`{e}r, and Katherine Lee.
\newblock Scalable extraction of training data from (production) language
  models.
\newblock \emph{arXiv preprint arXiv:2311.17035}, 2023.

\bibitem[Nicholson et~al.(2021)]{nicholson2021scite}
Joshua~M. Nicholson et~al.
\newblock scite: A smart citation index.
\newblock \emph{Quantitative Science Studies}, 2\penalty0 (3):\penalty0
  882--898, 2021.
\newblock \doi{10.1162/qss_a_00146}.

\bibitem[Novoselov et~al.(2004)Novoselov, Geim, Morozov, Jiang, Zhang, Dubonos,
  Grigorieva, and Firsov]{novoselov2004}
K.~S. Novoselov, A.~K. Geim, S.~V. Morozov, D.~Jiang, Y.~Zhang, S.~V. Dubonos,
  I.~V. Grigorieva, and A.~A. Firsov.
\newblock Electric field effect in atomically thin carbon films.
\newblock \emph{Science}, 306\penalty0 (5696):\penalty0 666--669, 2004.
\newblock \doi{10.1126/science.1102896}.

\bibitem[Nuijten et~al.(2016)]{nuijten2016statcheck}
Mich{\`e}le~B. Nuijten et~al.
\newblock Statistical reporting errors in psychology.
\newblock \emph{Behavior Research Methods}, 48:\penalty0 1205--1226, 2016.
\newblock \doi{10.3758/s13428-015-0664-2}.

\bibitem[Page et~al.(1999)Page, Brin, Motwani, and Winograd]{pagerank1999}
Lawrence Page, Sergey Brin, Rajeev Motwani, and Terry Winograd.
\newblock The {PageRank} citation ranking: Bringing order to the web.
\newblock In \emph{Proceedings of the 7th International World Wide Web
  Conference}, 1999.
\newblock URL \url{http://ilpubs.stanford.edu:8090/422/1/1999-66.pdf}.

\bibitem[Russinovich(2025)]{russinovich2025refchecker}
Mark Russinovich.
\newblock {RefChecker}.
\newblock \url{https://github.com/markrussinovich/refchecker}, 2025.

\bibitem[Samsonau(2025)]{samsonau2025scicodelint}
Sergey Samsonau.
\newblock scicode-lint: Detecting methodology bugs in scientific {Python} code
  with {LLM}-generated patterns.
\newblock \emph{arXiv preprint arXiv:2603.17893}, 2025.

\bibitem[{Seeds of Science}(2024)]{fidx2024}
{Seeds of Science}.
\newblock Is a qualitative metric of falsifiability possible? {The} {F-index}.
\newblock \emph{Seeds of Science}, 2024.
\newblock \doi{10.53975/1y7h-g9wd}.

\bibitem[Sharma et~al.(2024)Sharma, Tong, et~al.]{sharma2024sycophancy}
Mrinank Sharma, Meg Tong, et~al.
\newblock Towards understanding sycophancy in language models.
\newblock In \emph{Proceedings of the Twelfth International Conference on
  Learning Representations (ICLR)}, 2024.

\bibitem[Simkin and Roychowdhury(2003)]{simkin2003read}
Mikhail~V. Simkin and Vwani~P. Roychowdhury.
\newblock Read before you cite!
\newblock \emph{Complex Systems}, 14:\penalty0 269--274, 2003.
\newblock \doi{10.25088/complexsystems.14.3.269}.

\bibitem[Sumner et~al.(2022)Sumner, Vitale, and McIntosh]{ripeta2021}
Josh~Q. Sumner, Cynthia~Hudson Vitale, and Leslie~D. McIntosh.
\newblock {RipetaScore}: Measuring the quality, transparency, and
  trustworthiness of a scientific work.
\newblock \emph{Frontiers in Research Metrics and Analytics}, 6:\penalty0
  751734, 2022.
\newblock \doi{10.3389/frma.2021.751734}.

\bibitem[Tahamtan and Bornmann(2019)]{tahamtan2019citingbehavior}
Iman Tahamtan and Lutz Bornmann.
\newblock What do citation counts measure? {An} updated review of studies on
  citations in scientific documents published between 2006 and 2018.
\newblock \emph{Scientometrics}, 121\penalty0 (3):\penalty0 1635--1684, 2019.
\newblock \doi{10.1007/s11192-019-03243-4}.

\bibitem[Tang et~al.(2024)]{tang2024minicheck}
Liyan Tang et~al.
\newblock {MiniCheck}: Efficient fact-checking of {LLMs}.
\newblock In \emph{Proceedings of the 2024 Conference on Empirical Methods in
  Natural Language Processing (EMNLP)}, pages 8818--8847, 2024.
\newblock \doi{10.18653/v1/2024.emnlp-main.499}.

\bibitem[Thagard(1989)]{thagard1989explanatory}
Paul Thagard.
\newblock Explanatory coherence.
\newblock \emph{Behavioral and Brain Sciences}, 12\penalty0 (3):\penalty0
  435--502, 1989.
\newblock \doi{10.1017/S0140525X00057046}.

\bibitem[Wadden et~al.(2020)]{wadden2020scifact}
David Wadden et~al.
\newblock Fact or fiction: Verifying scientific claims.
\newblock In \emph{Proceedings of the 2020 Conference on Empirical Methods in
  Natural Language Processing (EMNLP)}, pages 7534--7550, 2020.
\newblock \doi{10.18653/v1/2020.emnlp-main.609}.

\bibitem[Waltman and van Eck(2012)]{waltman2012hindex}
Ludo Waltman and Nees~Jan van Eck.
\newblock The inconsistency of the h-index.
\newblock \emph{Journal of the American Society for Information Science and
  Technology}, 63\penalty0 (2):\penalty0 406--415, 2012.
\newblock \doi{10.1002/asi.21678}.

\bibitem[Wang et~al.(2025)]{wang2025novelty}
Yutao Wang et~al.
\newblock A review on the novelty measurements of academic papers.
\newblock \emph{arXiv preprint arXiv:2501.17456}, 2025.

\bibitem[Wei et~al.(2023)Wei, Kim, Tay, and Le]{wei2023ushaped}
Jason Wei, Najoung Kim, Yi~Tay, and Quoc~V. Le.
\newblock Inverse scaling can become {U}-shaped.
\newblock In \emph{Proceedings of the 2023 Conference on Empirical Methods in
  Natural Language Processing (EMNLP)}, pages 15580--15591, 2023.
\newblock \doi{10.18653/v1/2023.emnlp-main.963}.
\newblock arXiv:2211.02011.

\bibitem[Wilsdon et~al.(2015)Wilsdon, Allen, Belfiore, Campbell, Curry, Hill,
  Jones, Kain, Kerridge, Thelwall, Tinkler, Viney, Wouters, Hill, and
  Johnson]{wilsdon2015metrictide}
James Wilsdon, Liz Allen, Eleonora Belfiore, Philip Campbell, Stephen Curry,
  Steven Hill, Richard Jones, Roger Kain, Simon Kerridge, Mike Thelwall, Jane
  Tinkler, Ian Viney, Paul Wouters, Jude Hill, and Ben Johnson.
\newblock The {Metric Tide}: Report of the independent review of the role of
  metrics in research assessment and management.
\newblock Technical report, Higher Education Funding Council for England
  ({HEFCE}), 2015.

\bibitem[Xu et~al.(2026)Xu, Qiu, Sun, et~al.]{xu2026ghostcite}
Zuyao Xu, Yuqi Qiu, Lu~Sun, et~al.
\newblock {GhostCite}: A large-scale analysis of citation validity in the age
  of large language models.
\newblock \emph{arXiv preprint arXiv:2602.06718}, 2026.

\bibitem[Yuan et~al.(2026)Yuan, Shi, Zhang, Sun, Chawla, and
  Ye]{yuan2026citeaudit}
Zhengqing Yuan, Kaiwen Shi, Zheyuan Zhang, Lichao Sun, Nitesh~V. Chawla, and
  Yanfang Ye.
\newblock {CiteAudit}: You cited it, but did you read it? {A} benchmark for
  verifying scientific references in the {LLM} era.
\newblock \emph{arXiv preprint arXiv:2602.23452}, 2026.
\newblock v1 accessed via Wayback; withdrawn from arXiv 2026-04-27.

\bibitem[Zha et~al.(2023)]{zha2023alignscore}
Yuheng Zha et~al.
\newblock {AlignScore}: Evaluating factual consistency with a unified alignment
  function.
\newblock In \emph{Proceedings of the 61st Annual Meeting of the Association
  for Computational Linguistics (ACL)}, pages 11328--11348, 2023.
\newblock \doi{10.18653/v1/2023.acl-long.634}.

\end{thebibliography}

\end{document}